\documentclass[aps,prd,reprint,notitlepage,onecolumn,groupedaddress]{revtex4-1}
\usepackage{amsmath}
\usepackage{amsfonts}
\usepackage{amssymb}
\usepackage{textcomp}

\newcommand{\etal}{\textit{et al}.}

\begin{document}
\title{Inflation with an antisymmetric tensor field}

\author{Sandeep Aashish}
\email[]{sandeepa16@iiserb.ac.in}

\author{Abhilash Padhy}
\email[]{abhilash92@iiserb.ac.in}

\author{Sukanta Panda}
\email[]{sukanta@iiserb.ac.in}

\author{Arun Rana}
\email[]{arunrana@iiserb.ac.in}

\affiliation{Department of Physics, Indian Institute of Science Education and Research, Bhopal 462066, India}

\date{\today}

\begin{abstract}
We investigate the possibility of inflation with models of antisymmetric tensor field having minimal and nonminimal couplings to gravity. Although the minimal model does not support inflation, the nonminimal models, through the introduction of a nonminimal coupling to gravity, can give rise to stable de-Sitter solutions with a bound on the coupling parameters. The values of field and coupling parameters are sub-planckian. Slow roll analysis is performed and slow-roll parameters are defined which can give the required number of e-folds for sufficient inflation. Stability analysis has been performed for perturbations to antisymmetric field while keeping the metric unperturbed, and it is found that only the sub-horizon modes are free of ghost instability for de-Sitter space.
\end{abstract}
\maketitle

\section{Introduction}

Inflation as a theory, has been successfull in describing the structure and evolution of our universe \cite{guth1981,starobinsky1980}. As ordinary matter or radiation can not source inflation, several models have been built to describe inflation where a hypothetical field may it be scalar, vector or tensor drives the inflation \cite{martin2014}. Many theories have considered the scalar field called ``inflaton" as the source for inflation and are able to describe the cosmology of universe \cite{gottlober1990,roberts1994,parsons1995,barrow1995,linde1986,linde1993}. Most of the scalar field models having simple form of potential are ruled out as they are not compatible with the Planck's observational data for the cosmic microwave background \cite{martin2014,yuan2011,gomes2018}. Another class of models considers a vector field as an alternative to the inflaton \cite{ford1989,burd1991,golovnev2008,darabi2014,bertolami2015}. But almost all of these models suffer from instabilities like ghost instability \cite{peloso2009} and gradient instability \cite{ryonamba2017} which leads to an unstable vacuum. 
       
As the quantum corrections in cosmology and their possible phenomenological implications are becoming relevant \cite{fabris2012}, models with connections to high energy theories like the string theories provide an interesting alternative to traditional inflation model building. A particular theory of interest is that of a rank-2 antisymmetric tensor field, which appears in all superstring models  \cite{rohm1986,ghezelbash2009}. Antisymmetric tensors have been studied before in several aspects, including phase transitions, strong-weak coupling duality  \cite{quevedo1996,olive1995,polchinski1995,siegel1980,hata1981,buchbinder1988,duff1980,bastianelli2005a,*bastianelli2005b} and even some astrophysical aspects \cite{damour1994}. More recently, quantum aspects of antisymmetric fields in different settings have been studied   \cite{altschul2010,buchbinder2008,shapiro2016,aashish2018,aashish2018b}. However, efforts for cosmological studies with antisymmetric tensors were rare until the past decade. A few pertinent works with regard to inflation scenarios with $N$-form fields in anisotropic spacetime was carried out in Refs. \cite{koivisto2009a,koivisto2009b} and near a Schwarzschild metric in Ref. \cite{prokopec2006}. More recently, two-form fields have been studied in the context of anisotropic inflation \cite{asuka2015} and gravitational waves \cite{obata2018}. 

In this paper, we study the possibility of inflation with antisymmetric tensor field by considering minimal and nonminimal models originally considered in Altschul \etal  \cite{altschul2010}. We find that the minimal model does not support inflation. However, introducing a new parameter in the form of nonminimal coupling to gravity helps to achieve inflation. The nonminimal coupling terms we incorporate here are part of a general action constructed in \cite{altschul2010} and are inspired by spontaneous Lorentz violation theories. The most general nonminimal nonderivative couplings upto quadratic order in antisymmetric tensor $B_{\mu\nu}$ (restricted to parity-even terms) are written as \cite{altschul2010}
\begin{eqnarray}
\mathcal{L}_{NM}=\frac{1}{2\kappa}\xi B^{\mu \nu} B_{\mu \nu} R + \frac{1}{2 \kappa} \zeta B^{\lambda \nu} B^{\mu}_{\ \nu} R_{\lambda \mu} + \frac{1}{2 \kappa}\gamma B^{\kappa\lambda}B^{\mu\nu}R_{\kappa\lambda\mu\nu}
\end{eqnarray}
Demanding a stable Schwarzschild solution, we do not consider the coupling with $R_{\kappa\lambda\mu\nu}$, but we will consider the remaining couplings ($\xi$ and $\zeta$ term) because our model does not contain the cosmological constant ($\Lambda$) \cite{prokopec2006}. 
We also set up a perfect slow roll scenario for this inflationary model, prior to developing a full perturbation theory for antisymmetric tensor in future works. However, an instability analysis for the perturbations to only the antisymmetric tensor field is performed. Although, in Ref. \cite{koivisto2009a} a similar instability analysis was done for $R$ coupling and possibility of ghosts was found, the present analysis is different in the following ways: (i) the spacetime is isotropic and homogeneous; (ii) background structure of $B_{\mu\nu}$ is specified; and (iii) choice of parameter space takes into account the conditions for slow-rolling inflation.
        
This work is organized as follows. In Sec. \ref{sec2}, we introduce background structures of the metric and the antisymmetric tensor, and establish the general setup of our analysis through a simple model of a massive antisymmetric tensor field minimally coupled to gravity. It is shown that minimal model cannot give rise to inflation. Three cases of nonminimally coupled extensions of this model are considered in Sec. \ref{sec3}. The conditions for inflation and the de-Sitter space solutions have been derived. In Sec. \ref{sec4}, we check the stability of possible de-Sitter space. In Sec. \ref{sec5}, the slow roll parameters for the nonminimal models are constructed and the number of e-folds are calculated. Sec. \ref{sec6} presents stability analysis for perturbations to antisymmetric tensor field, while keeping the metric unperturbed.

\section{\label{sec2}Minimal model and the setup}
\subsection{Setup}
As a first step towards studying an inflation model and as a precursor to extracting phenomenological results like the power spectrum, which come from the dynamics of perturbations to background fields (the metric and inflation field), it makes sense to establish a theory of background fields that ensures  \cite{dodelson2003,weinberg2008}:
\begin{enumerate}
\item a de-Sitter space solution exists,
\item the de-Sitter space should be stable, i.e. perturbations to solutions must decay with time, and
\item more than 70 efolds of slow-roll inflation.
\end{enumerate}

An obvious choice for the background metric is the Friedmann–Lemaître–Robertson–Walker (FLRW) metric, motivated by the cosmological principle that imposes homogeneity and isotropy symmetries on the background universe. With the choice of metric signature ($-+++$), the (background) metric components $g_{\mu\nu}$ read,
 \begin{equation}
 \label{amms1}
 g_{00} = -1,  \quad  g_{ij} = a(t)^2 \delta_{ij},
 \end{equation}
where $a(t)$ is the scale factor for expansion of universe. The Riemann–Christoffel tensor, Ricci tensor and Ricci scalar in terms of metric components in Eq. (\ref{amms1}) are given by,
 \begin{equation}
 \label{amms2}
  R_{0i0j} = - a\ddot{a}\delta_{ij},  \quad \quad 
  R_{ijkl} = \delta_{ik} \delta_{jl} (a\dot{a})^2 \quad i<j;
 \end{equation}
 
 \begin{equation}
 \label{amms3}
  R_{00} = - 3\frac{\ddot{a}}{a}, \quad  \quad R_{ij} = \delta_{ij}(a\ddot{a} + 2\dot{a}^2);
 \end{equation}
 
 \begin{equation}
 \label{amms4}
 R = 6\Big[  \frac{\ddot{a}}{a} + \Big(\frac{\dot{a}}{a}  \Big)^2 \Big].
 \end{equation}
 
We are interested in a theory where the inflation-driving field is an antisymmetric tensor $B_{\mu \nu}$,
    \begin{equation}
    \label{amms5}
      B_{\mu \nu} = - B_{\nu \mu}.
    \end{equation}
In general, $B_{\mu\nu}$ has six independent components and a structure similar to that of the electromagnetic field strength tensor. A convenient representation of $B_{\mu\nu}$, analogous to the electrodynamic decomposition of field strength into electric and magnetic fields, is given by  \cite{altschul2010}, 
\begin{eqnarray}
\label{amms6}
B_{0j}=-\Sigma^{j}, \quad B_{jk}=\epsilon_{jkl}\Xi^{l}.
\end{eqnarray}
An interesting but also challenging aspect of cosmology with antisymmetric tensors is that the perturbations to all six components will play a role in the dynamics, and could offer important phenomenology. However, for setting up the background dynamics, we can exploit the freedom to choose a structure for $B_{\mu\nu}$ that simplifies the calculations of the present work without losing generality. As will be seen shortly, this choice of $B_{\mu\nu}$ structure manifests in the constraint equations for off-diagonal components of spatial part of energy-momentum tensor, ensuring homogeneity and isotropy of background metric $g_{\mu\nu}$. For our convenience, we choose $\Sigma^{j}=0$, and $\Xi^{l}=B(t)$, $l=1,2,3$, so that,
 \begin{equation}
 \label{bmms1}
  B_{\mu \nu} =  \begin{pmatrix}
 0 & 0 & 0 & 0 \\
 0 & 0 & B(t) & -B(t) \\
 0 & -B(t) & 0 & B(t) \\
 0 & B(t) & -B(t) & 0
 \end{pmatrix}.
 \end{equation}
 
\subsection{Minimal model}
At this point, to set up our approach, we consider a ``minimal" model of an antisymmetric tensor first considered in Ref. \cite{altschul2010},
\begin{equation}
\label{bmms2}
S = \int d^4x \sqrt{-g} \Big[ \frac{R}{2\kappa} - \frac{1}{12} H_{\lambda\mu\nu}(B) H^{\lambda\mu\nu}(B) - V(B)\Big],
\end{equation}
where $H_{\lambda \mu \nu}  = \nabla_{\lambda} B_{\mu \nu} + \nabla_{\mu} B_{\nu \lambda} +  \nabla_{\nu} B_{\lambda \mu}$ is the gauge-invariant kinetic term \cite{altschul2010} ($\nabla_{\mu}$ is the covariant derivative), and $V(B)$ is the potential term. Rest of the symbols have their usual meanings, with $g$ being the metric determinant, $R$ the Ricci scalar and $\kappa$ the inverse square of Planck mass $M_{pl}$. For the present problem, we will only consider quadratic potential of the form $m^{2}B_{\mu\nu}B^{\mu\nu}/4$, though some of the expressions (especially for slow roll analysis) will be written in terms of $V(B)$ for generality. 

Here onwards, we omit the arguments of functions and functionals ($a(t), B(t), V(B)$, etc.) for notational convenience and their functional dependence is assumed until stated otherwise. 

Our starting point for finding de-Sitter space solutions is the Einstein equation, obtained by varying the action (\ref{bmms2}) with respect to metric $g_{\mu\nu}$, 
\begin{eqnarray}
\label{bmms3}
G_{\mu\nu}=\kappa T^{M}_{\mu\nu},
\end{eqnarray}
where, $G_{\mu\nu}$ is the Einstein tensor and the energy momentum tensor $T^{M}_{\mu\nu}$ is given by, 
\begin{eqnarray}
\label{dmms1}
T^{M}_{\mu\nu}=\dfrac{1}{2}H^{\alpha \beta}_{\quad \mu} H_{\nu \alpha \beta}+ m^2 B^{\alpha}_{\ \mu} B_{\alpha \nu} - g_{\mu \nu} (\frac{1}{12} H_{\alpha \beta \gamma} H^{\alpha \beta \gamma} + \frac{1}{4}m^2 B_{\alpha \beta}B^{\alpha \beta}).
\end{eqnarray}
It can be inferred from Eq. (\ref{dmms1}) that $T^{M}_{\mu\nu}$ in general has off-diagonal elements. One can always choose a structure for $B_{\mu\nu}$ that renders the off-diagonal elements of the spatial components of EM tensor, $T^{M}_{ij}$, equal to zero albeit with a caveat that the pressure ($T^{M}_{ii}$) becomes anisotropic, i.e. $T^{M}_{11}\neq T^{M}_{22} \neq T^{M}_{33}$. For our choice of $B_{\mu\nu}$, Eq. (\ref{bmms1}), the isotropy of pressure is ensured while introducing an additional constraint on the off-diagonal components $T^{M}_{ij}$.

We define, 
\begin{equation}
 \label{cmms4}
 B(t) = a(t)^2 \phi(t),
 \end{equation}
so as to obtain a familiar form of equations of motion, resembling that of scalar field models. Choosing the quadratic potential, $V(B) = m^{2}B_{\mu\nu}B^{\mu\nu}/4 = 3m^{2}\phi^{2}/2$ in the background FRW metric Eq. (\ref{amms1}) and the background tensor field Eq. (\ref{bmms1}), Eq. (\ref{bmms3}) takes the form,
\begin{eqnarray}
\label{bmms4}
G_{00}=T^{M}_{00} &\implies & H^2 = \frac{\kappa}{2}[(\dot{\phi} + 2 H\phi)^2 +m^{2}\phi^{2}], \\
\label{bmms5}
G_{ij}=T^{M}_{ij} &\implies & 2\dot{H} + 3H^{2} = \frac{\kappa}{2}[(\dot{\phi} + 2 H\phi)^2 - m^{2}\phi^{2}], \quad i=j,
\end{eqnarray}
As pointed out before, in addition to Eqs. (\ref{bmms4}) and (\ref{bmms5}), the off-diagonal components $T^{M}_{ij}$ ($i\neq j$) satisfy the following constraint equation:
\begin{eqnarray}
\label{bmms6}
\frac{\kappa}{2}[(\dot{\phi} + 2 H\phi)^2 - m^{2}\phi^{2}] = 0, 
\end{eqnarray}
ensuring that the symmetries of spacetime (homogeneity and isotropy) are maintained.
The equation of motion for $\phi$ can be obtained from the energy-momentum conservation equation $\nabla^{\mu}T^{M}_{\mu\nu} = 0$, but we do not write it here explicitly because it is not an independent equation and hence is irrelevant for the current calculations. 
Using the constraint Eq. (\ref{bmms6}) in Eq. (\ref{bmms5}), we obtain
\begin{eqnarray}
\label{bmms7}
  \frac{\ddot{a}}{a} = -\dfrac{H^{2}}{2}.
\end{eqnarray}
Clearly, the acceleration of $a(t)$ is negative and hence the minimal model does not support the possibility of inflation. Eq. (\ref{bmms7}) provides an insight into what modifications could be made to the action (\ref{bmms2}) to allow inflation. A straightforward solution for positive acceleration would be to incorporate additional parameter in the rhs of Eq. (\ref{bmms7}) such that $\ddot{a}$ has nontrivial solutions. In subsequent sections, we consider an extension of this model consisting of nonminimal coupling of $B_{\mu\nu}$ with gravity that resolves this issue.

\section{\label{sec3}Nonminimal models} 
\subsection{The models}
The requirement of positive acceleration of the scale factor is met by a simple extension of theory (\ref{bmms2}) consisting of a nonminimal coupling to gravity  \cite{altschul2010} given by:
\begin{eqnarray}
\label{anon1}
S = \int d^{4}x \sqrt{-g} \Big[ \frac{R}{2\kappa} - \frac{1}{12} H_{\lambda \mu \nu} H^{\lambda \mu \nu} 
- \dfrac{m^{2}}{4} B_{\mu\nu}B^{\mu\nu} + \mathcal{L}_{NM}\Big] ,
\end{eqnarray}
where $\mathcal{L}_{NM}$ is the non-minimal coupling term. As mentioned before, we will consider two cases, with $\mathcal{L}_{NM}=\frac{1}{2\kappa}\xi B^{\mu \nu} B_{\mu \nu} R$ and $\mathcal{L}_{NM} =  \frac{1}{2 \kappa}\zeta B^{\lambda \nu} B^{\mu}_{\ \nu} R_{\lambda \mu}$ separately, for convenience. 
The non-minimal coupling term $\mathcal{L}_{NM}$, is parametrized by $\xi$ and $\zeta$ for couplings with $R$ and $R_{\mu\nu}$ respectively. The parameters $\xi$ and $\zeta$ have dimensions of $M_{pl}^{-2}$.

\subsubsection{Case: $\mathcal{L}_{NM}=\frac{1}{2\kappa}\xi B^{\mu \nu} B_{\mu \nu} R$}
With $\mathcal{L}_{NM}=\frac{1}{2\kappa}\xi B^{\mu \nu} B_{\mu \nu} R$ in Eq. (\ref{anon1}), the corresponding energy momentum tensor is given by,
\begin{eqnarray}
\label{cnon1}
T_{\mu\nu} = T_{\mu\nu}^{M} + T_{\mu\nu}^{\xi},
\end{eqnarray}
where, 
\begin{eqnarray}
\label{cnon2}
T_{\mu\nu}^{\xi} = \dfrac{\xi}{\kappa}\left[\nabla_{\mu} \nabla_{\nu}(B_{\alpha \beta}B^{\alpha \beta}) - g_{\mu \nu} 
\nabla^{\lambda} \nabla_{\lambda} (B_{\alpha \beta}B^{\alpha \beta}) - G_{\mu \nu} (B_{\alpha \beta}B^{\alpha \beta}) - 2R  B^{\alpha}_{\ \mu} B_{\alpha \nu}   \right].
\end{eqnarray}
Following the steps of previous section, we write the Einstein equations,
\begin{eqnarray}
\label{bnon1}
G_{00}= \kappa T_{00} &\implies & H^2 + 6\xi (2H\phi \dot{\phi} + H^2 \phi^2) = \frac{\kappa}{2}[(\dot{\phi} + 2 H\phi)^2 +m^{2}\phi^{2}], \\
\label{bnon2}
G_{ij}=\kappa T_{ij} &\implies & 2\dot{H} + 3H^{2} + 6\xi (2\phi \ddot{\phi} + 2\dot{\phi}^2 - 2\dot{H}\phi^2 - 5 H^2\phi^2 + 4H\phi \dot{\phi} ) \nonumber \\ && = \frac{\kappa}{2}[(\dot{\phi} + 2 H\phi)^2 - m^{2}\phi^{2}], \quad i=j, 
\end{eqnarray}
Similarly, the constraint equation for off-diagonal components of $T_{ij}$ becomes,
\begin{eqnarray}
\label{bnon3}
\frac{\kappa}{2}[(\dot{\phi} + 2 H\phi)^2 - m^{2}\phi^{2}] = -6\xi(\dot{H} + 2H^{2})\phi^{2}.
\end{eqnarray}

\subsubsection{Case: $\mathcal{L}_{NM} = \frac{1}{2 \kappa} \zeta B^{\lambda \nu} B^{\mu}_{\ \nu} R_{\lambda \mu}$}
Substituting $R_{\mu \nu}$ coupling term, $\mathcal{L}_{NM} =  \zeta \frac{\sqrt{-g}}{2 \kappa} B^{\lambda \nu} B^{\mu}_{\ \nu} R_{\lambda \mu}$, in the action (\ref{anon1}), the energy momentum tensor in this case is given by,
\begin{eqnarray}
\label{nbon1}
T_{\mu\nu} = T_{\mu\nu}^{M} + T_{\mu\nu}^{\zeta},
\end{eqnarray}
where,
 \begin{eqnarray}
 \label{nbon2}
 T^{\zeta}_{\mu \nu} = \frac{\zeta}{\kappa} \Big[ \frac{1}{2} g_{\mu \nu} (B^{\alpha \gamma} B^{\beta}_{\ \gamma}R_{\alpha \beta} - \nabla_{\alpha} \nabla_{\beta} B^{\alpha \gamma} B^{\beta}_{\ \gamma}) - B^{\alpha }_{\ \mu} B^{\beta}_{\ \nu}R_{\alpha \beta}
  -B^{\alpha \beta} B_{\mu \beta}R_{ \nu \alpha} - B^{\alpha \beta} B_{\nu \beta}R_{ \mu \alpha}&& \nonumber \\ 
 + \frac{1}{2}(\nabla_{\alpha} \nabla_{\mu} B_{\nu \beta} B^{\alpha \beta} +\nabla_{\alpha} \nabla_{\nu} B_{\mu \beta} B^{\alpha \beta} - \nabla^{\lambda} \nabla_{\lambda} B^{\alpha}_{\ \mu} B_{\alpha \nu})  \Big],
 \end{eqnarray}
Similarly, the Einstein equations are found to be 
 \begin{eqnarray}
\label{nbnon3}
G_{00}= \kappa T_{00} &\implies & H^2 + 2\zeta H\phi \dot{\phi}  = \frac{\kappa}{2}[(\dot{\phi} + 2 H\phi)^2 +m^{2}\phi^{2}], \\
\label{nbnon4}
G_{ij}=\kappa T_{ij} &\implies & 2\dot{H} + 3H^{2} + \zeta (2\phi \ddot{\phi} + 2\dot{\phi}^2 - 4\dot{H}\phi^2 - 12 H^2\phi^2  ) \nonumber \\ && = \frac{\kappa}{2}[(\dot{\phi} + 2 H\phi)^2 - m^{2}\phi^{2}], \quad i=j, 
\end{eqnarray}
For the off-diagonal components, the constraint equation reads,
\begin{equation}
\label{nbnon5}
\frac{\kappa}{2}[(\dot{\phi} + 2 H\phi)^2 - m^{2}\phi^{2}] = -\zeta(\dot{H} + 3H^{2})\phi^{2}+ \frac{\zeta}{2}(\phi \ddot{\phi} + \dot{\phi}^2 + 3H\phi \dot{\phi}), \quad i\neq j
\end{equation}
\subsection{de-Sitter solutions}
To find the de-Sitter solutions, we consider the fact that an exponential expansion of the universe (during inflation) implies a constant Hubble parameter, $H=H_{0}$. Moreover, it helps to further take into account the slow rolling of field $\phi$ during inflation, so that it can be thought of as nearly constant, $\phi\approx\phi_{0}$. The question of whether an exact de-Sitter space exists boils down to finding non-zero solutions ($\phi_{0},H_{0}$) to the Einstein equations (\ref{bnon1}) $-$ (\ref{bnon3}) in the de-Sitter limit, $ \dot{H} = \dot{\phi} = 0$. 
First, using the constraint Eq. (\ref{bnon3}) in Eq. (\ref{bnon2}), we get for the coupling with $R$,
\begin{eqnarray}
\label{bnon4}
2\dot{H} + 3H^{2} + 12\xi (\phi \ddot{\phi} + \dot{\phi}^2  + 2H\phi \dot{\phi} - \frac{1}{2}\dot{H}\phi^2 - \frac{3}{2} H^2\phi^2 ) = 0. 
\end{eqnarray}
Then, applying the de-Sitter limit to Eqs. (\ref{bnon1}) and (\ref{bnon4}), de-Sitter solutions $\phi_0$ and $H_0$ are obtained. The results, including a similar calculation for the $R_{\mu\nu}$ coupling, are given in Table \ref{tab1} above.
\begin{table}[h!]
  \begin{center}
    \begin{tabular}{|c|c|c|c|} 
    \hline
      $\mathcal{L}_{NM}$ & \ $\phi_{0}^{2}$ \ & $H_{0}^{2}$ & Condition\\
      \hline
      $\dfrac{1}{2\kappa}\xi B^{\mu \nu} B_{\mu \nu} R$ & $\dfrac{1}{6\xi}$ & $\dfrac{\kappa m^{2}}{4(6\xi - \kappa)}$ & $\xi > \dfrac{\kappa}{6}$\\
      \hline
      $\dfrac{1}{2 \kappa} \zeta B^{\lambda \nu} B^{\mu}_{\ \nu} R_{\lambda \mu}$ & $\dfrac{1}{3\zeta}$ & $\dfrac{\kappa m^2}{2(3 \zeta - 2 \kappa)}$ & $\zeta > \dfrac{2 \kappa}{3}$\\
      \hline
    \end{tabular}
    \caption{The de-Sitter space solutions of $\phi$ and $H$, along with the condition on parameters $\xi$ and $\zeta$ corresponding to $R$ and $R_{\mu\nu}$ coupling terms respectively.}
    \label{tab1}
  \end{center}
\end{table}

It is worth noting that value of $\phi$ is sub-planckian in both cases. An interesting observation in the context of theories (\ref{bmms2}) and (\ref{anon1}) is that adding a nonminimal coupling gives rise to de-Sitter solutions, which are otherwise absent in minimal model. This is a unique feature of antisymmetric field models in contrast to the nonminimal models of scalar field inflation (see  \cite{nakanishi2018} and references therein). 
   

\section{\label{sec4}Stability analysis of the De-Sitter background}
In this section, the dynamics of nonminimal model (\ref{anon1}) around the de-Sitter background is analyzed. Whether a stable de-Sitter background is possible or not, can be checked by perturbing the field $\phi(t)$ and Hubble parameter $H(t)$ about de-Sitter solutions $H_{0}$ and $\phi_{0}$. The condition for stability is that the perturbations $\delta\phi (t)$ and $\delta H (t)$ must decay over time. The corresponding perturbations are given by,     
      \begin{equation}
      \label{asta1}
      H = H_0 + \delta H; \quad \quad \phi = \phi_0 + \delta \phi,
\end{equation}       
where $\delta H$ and $\delta \phi$ are small perturbation about $(H_0,\phi_0)$ in linear order. Substituting Eq. (\ref{asta1}) in Eqs. (\ref{bnon1}) and (\ref{bnon4}), and using the results in Table \ref{tab1}, we obtain, upto first order in perturbations,
\begin{eqnarray}
\label{asta4}
\dot{\delta \phi} &=& \dfrac{6\xi}{6\xi - \kappa} H_0 \delta \phi -  2 \phi_0 \delta H, \\
\label{asta5}
\dot{\delta H} &=& \left(\dfrac{12\xi(8\xi - \kappa)}{(6\xi - \kappa)^{2}}\right)\kappa H_{0}^{2}\phi_{0}\delta\phi - \dfrac{8}{3}\left(\dfrac{9\xi - \kappa}{6\xi - \kappa}\right)H_{0}\delta H.
\end{eqnarray}
Eq.(\ref{asta4}) and Eq.(\ref{asta5}) can be together expressed in the matrix form as follows:
\begin{equation}
\label{asta6}
\dot{\Theta} = A \Theta,
\end{equation}
where $\Theta$ is a column matrix and $A$ is a $(2\times 2)$ square matrix, given by,

\begin{eqnarray}
\label{asta7}
\Theta = \begin{pmatrix}
\delta \phi \\
\delta H
\end{pmatrix};
\quad \quad
A = \begin{pmatrix}
 \dfrac{6\xi}{6\xi - \kappa} H_0 & -  2 \phi_0 \\
 \left(\dfrac{12\xi(8\xi - \kappa)}{(6\xi - \kappa)^{2}}\right)\kappa H_{0}^{2}\phi_{0}\ & - \dfrac{8}{3}\left(\dfrac{9\xi - \kappa}{6\xi - \kappa}\right)H_{0}
\end{pmatrix}.
\end{eqnarray}
Upon solving Eq. (\ref{asta6}) the solution for $\Theta (t)$ has a general form,
\begin{eqnarray}
\label{csta0}
\Theta (t) = A_{1}e^{\lambda_{1}t} + A_{2}e^{\lambda_{2}t}.
\end{eqnarray}
The eigenvalues  $\lambda_1$ and $\lambda_2$ of the matrix $A$ can be calculated from its trace and determinant, which are
 \begin{eqnarray}
 \label{asta8}
 Tr[A] &=& \lambda_1 + \lambda_2 = -H_{0}\left(3 + \dfrac{\kappa}{3(6\xi - \kappa)}\right) \equiv -\tau H_{0},\\
 \label{asta9}
 \det[A] &=& \lambda_1 \lambda_2
 = - 4H_{0}^{2}.
  \end{eqnarray}
From Eqs. (\ref{asta8}) and (\ref{asta9}), one can deduce that one of the $\lambda_1$ and $\lambda_2$ is negative, and the negative eigenvalue strongly dominates the positive one. Moreover, due to the condition on $\xi$ as in Table \ref{tab1}, $\lambda_1 + \lambda_2 < -3H_{0}$. In view of Eq. (\ref{csta0}), it implies that the perturbations will grow exponentially over time due to the small positive eigenvalue, and thus will be unstable. However, it should still be possible to suppress this instability by constraining the coefficient of the growing part of $\Theta (t)$ in Eq. (\ref{csta0}), but it needs to be checked whether or how that can be achieved for the epoch of interest. In fact, considering the explicit solutions for $\lambda_{1,2}$,
\begin{eqnarray}
\label{bsta0}
\lambda_{1(2)} = -\dfrac{H_{0}}{2}[\tau - (+) \sqrt{\tau^{2} + 16}],
\end{eqnarray}
in the limiting case where $6\xi/\kappa \to 1^{+}$, $\lambda_{1}\to 0$ and $\lambda_{2}<< 0$. Substituting in Eq. (\ref{csta0}) leads to $\Theta (t) \approx A_{1} + A_{2}e^{\lambda_{2}t}$, and will provide a decaying solution if $A_{1}\approx 0$. Though, it is not clear at this time how such a solution can be obtained without heavily constraining the parameters ($\xi,\zeta$) and coefficients ($A_{1,2}$), thus we leave this problem for consideration in future.

\subsubsection{Case: $\mathcal{L}_{NM} = \frac{1}{2 \kappa} \zeta B^{\lambda \nu} B^{\mu}_{\ \nu} R_{\lambda \mu}$}
A similar analysis for the second case, $\mathcal{L}_{NM} = \frac{1}{2 \kappa} \zeta B^{\lambda \nu} B^{\mu}_{\ \nu} R_{\lambda \mu}$, leads to the following structure for matrix $A$ of Eq. (\ref{asta6}),
\begin{eqnarray}
\label{bsta1}
A = \begin{pmatrix}
 \dfrac{3 \zeta}{\zeta - \kappa} H_0 &  \dfrac{ 2\kappa - 3 \zeta}{\zeta - \kappa} \phi_0 \\
 -9 \dfrac{(\zeta - 4\kappa)(2\zeta - \kappa)}{(\zeta - \kappa)} H_{0}^{2}\phi_{0}\ & - \dfrac{(2\zeta + \kappa)(3\zeta - 2\kappa)}{\zeta (\zeta - \kappa)}H_{0}
\end{pmatrix}.
\end{eqnarray}
  The eigen values are calculated to be 
  \begin{eqnarray}
  \label{bsta2}
  \lambda_{1(2)} = -\dfrac{3H_{0}}{2}\left\{-1 - \dfrac{2\kappa}{3\zeta}- (+) \sqrt{\Big[1+\dfrac{2\kappa}{3\zeta}\Big]^{2} + 16\Big[1-\dfrac{2\kappa}{3\zeta}\Big]}\right\},
  \end{eqnarray}
which are again similar to $\xi$ case, Eq. (\ref{bsta0}), in the sense that one of the eigenvalues dominates over the other. 
  
\section{\label{sec5}Slow roll parameters}
We now consider a nearly de-Sitter spacetime for building an inflationary model. For a successful inflation, the duration of inflation should be more than 70 efolds   \cite{dodelson2003}. Slow roll parameters are introduced in a theory to control (i) the acceleration of universe, and (ii) the duration of inflation. One of the slow roll conditions relevant for the acceleration is $\epsilon$, given in terms of Hubble parameter, 
\begin{equation}
\label{aslo3}
\epsilon = - \frac{\dot{H}}{H^2}.
\end{equation}
Eq. (\ref{aslo3}) can be rewritten as
\begin{eqnarray}
\label{bslo0}
\frac{\ddot{a}}{a} = H^2(1- \epsilon),
\end{eqnarray}
and it can be seen that $\epsilon$ has to be small in order for acceleration to be positive.
A second slow roll parameter in terms of $\phi$ must be introduced to control the duration of inflation. A standard approach is to choose a parameter such that the equations of motions can be expressed in terms of slow roll parameters, and a relation between the two parameters can be obtained. Slow roll condition is satisfied if the smallness of one parameter is compatible with that of the other. 

For any arbitrary potential the equations of motion for this model can be written as
   \begin{eqnarray}
   \label{aslo1}
H^2 + (6\xi - 2\kappa)H^2\phi^2 + (12\xi - 2\kappa)H\phi\dot{\phi} - \frac{\kappa}{2}\dot{\phi}^2 - \frac{\kappa V}{3} = 0, \\
\label{aslo2}
2\dot{H} + 3H^{2} + 12\xi (\phi \ddot{\phi} + \dot{\phi}^2  + 2H\phi \dot{\phi} - \frac{1}{2}\dot{H}\phi^2 - \frac{3}{2} H^2\phi^2 ) = 0. 
\end{eqnarray}
We now introduce a second slow roll parameter $\delta \equiv \dfrac{\dot{\phi}}{H\phi}$. Dividing by $H^{2}$, Eq. (\ref{aslo2}) can be expressed in terms of the new slow roll parameter $\delta$,
\begin{eqnarray}
\label{bslo1}
3-18\tau - 2\epsilon + 12\tau[\dfrac{\dot{\delta}}{H} + 2\delta^{2} + (2-\epsilon)\delta + \dfrac{\epsilon}{2}] = 0,
\end{eqnarray}
where, $\tau \equiv \xi\phi^{2}$. During inflation, we can take the value of $\tau$ to be of the same order as that in a de-Sitter spacetime, i.e. $\tau\approx 1/6$.  
In Eq. (\ref{aslo1}), using the slow roll condition $\dot{\phi}^{2}<V$ and taking its derivative, we obtain 
\begin{eqnarray}
\label{bslo3}
\epsilon = \delta\left[\dfrac{(6\xi - 2\kappa)\phi^{2}}{1 + (6\xi - 2\kappa)\phi^{2} + \delta(12\xi - 2\kappa)\phi^{2}} - \dfrac{\phi V_{\phi}}{2V}\right].
\end{eqnarray}
where $V_{\phi}=dV/d\phi$. An explicit relation between $\epsilon$ and $\delta$ is obtained by using the flat potential condition, $V_{\phi}<<V$. The results for the two cases of nonminimal couplings are given in Table \ref{tab2}.
\begin{table}[h!]
\centering
\begin{tabular}{|c|c|}
\hline
$\mathcal{L}_{NM}$ & \ $\epsilon$ \ \\
      \hline
      $\dfrac{1}{2\kappa}\xi B^{\mu \nu} B_{\mu \nu} R$ & $\epsilon \approx \dfrac{\delta}{(6\xi - 2\kappa)^{-1}\phi^{-2} + 1}\sim \delta $ \\
      \hline
      $\dfrac{1}{2 \kappa} \zeta B^{\lambda \nu} B^{\mu}_{\ \nu} R_{\lambda \mu}$ & $\epsilon \approx  \dfrac{\delta}{1- (2 \kappa \phi^2)^{-1} } \sim \delta$ \\
      \hline
\end{tabular}
\caption{Relation between the slow-roll parameters $\epsilon$ and $\delta$ for each case of nonminimal coupling.}  
\label{tab2}
\end{table}

The small  $\delta$ indicates that the background field should be nearly constant which eventually leads to flat potential satisfying the requirement of slow roll. It is evident from Table \ref{tab2} that in both cases of nonminimal coupling, small $\delta$ gives rise to small $\epsilon$, thereby allowing slow-roll inflation. The duration of inflation can be expressed by the number of e-folds. Before calculating the number of e-fold it is important to calculate $\dot{\delta}$.
  \begin{equation}
  \label{aslo8}
  \dot{\delta} = \frac{\ddot{\phi}}{H \phi} - \frac{\dot{\phi}^2}{H\phi^2} - \frac{\dot{\phi} \dot{H}}{\phi H^2} = H \delta (\epsilon - \delta).
\end{equation}   

During inflation $H$ is nearly constant which says that $\dot{\delta}$ is approximately zero or $\delta$ is nearly constant during inflation. Now, the number of e-folds can be calculated to be,
\begin{equation}
\label{aslo9}
N = \int_{t_i}^t H dt = \int_{\phi_i}^{\phi} d\phi \frac{H}{\dot{\phi}} = \frac{1}{\delta}  \int_{\phi_i}^{\phi}  \frac{d\phi}{\phi} = \frac{1}{\delta} ln\left(\frac{\phi}{\phi_i}\right).
\end{equation}
Clearly, it is feasible now to get 70 or more e-folds since $\delta$ is the only controlling parameter, and its smallness ensures sufficient duration of slow-rolling inflation. 

\section{\label{sec6}Stability of perturbations to $B_{\mu\nu}$}
Although this model is able to provide a stable de-sitter type inflation with a lightly tuned nonminimal coupling with curvature terms, it should be free from the instabilities in order to give a sustainable inflationary model. A complete stability analysis would include perturbations to $B_{\mu\nu}$ and the metric. However, as an initial check, we consider here only the perturbations to the background antisymmetric tensor field $B_{\mu\nu}$, leaving the metric unperturbed. A similar analysis for the $R$ coupling was performed in Ref. \cite{koivisto2009a} in anisotropic spacetime. In the present case, we consider both couplings, i.e. $\mathcal{L}_{NM}=\frac{1}{2\kappa}\xi B^{\mu \nu} B_{\mu \nu} R + \frac{1}{2 \kappa}\zeta B^{\lambda \nu} B^{\mu}_{\ \nu} R_{\lambda \mu}$, and the spacetime background is homogeneous and isotropic. The choice of background structure of $B_{\mu\nu}$ remains the same as in Eq. (\ref{bmms1}). The perturbed field is given by $B_{\mu \nu} + \delta B_{\mu\nu}$, where
     \begin{eqnarray}
     \label{aper1}
     \delta B_{0i} = - E_i , \quad \quad \delta B_{ij} = \epsilon_{ijk} M_k.
     \end{eqnarray}
Substituting this perturbation in the action (\ref{anon1}) results in the perturbed action containing terms upto quadratic order in perturbation. We are essentially interested in the terms of second order in perturbation, because these contain kinetic terms corresponding to perturbations $E_{i}$ and $M_{i}$. The second order part of the perturbed action reads, 
     \begin{eqnarray}
     \label{aper2}
     S_2 = \int d^4 x \ \Bigg[ \frac{1}{2 a}\left( \dot{\vec{M}}\cdot\dot{\vec{M}} + 2\dot{\vec{M}}\cdot(\vec{\nabla} \times \vec{E}) + (\vec{\nabla} \times \vec{E}) \cdot(\vec{\nabla} \times \vec{E})  \right) - 
     \frac{1}{2 a^3}(\vec{\nabla}\cdot\vec{M})^2 \nonumber\\
     + \left( \frac{m^2}{2} - \dfrac{(6\xi + 2\zeta)}{\kappa} \dot{H} - \dfrac{(12\xi + 3\zeta)}{\kappa} H^2 \right)a(\vec{E}\cdot\vec{E}) \nonumber \\ 
    - \left( \frac{m^2}{2} - \dfrac{(6\xi + \zeta)}{\kappa} \dot{H} - \dfrac{(12\xi + 3\zeta)}{\kappa} H^2 \right)\dfrac{(\vec{M}\cdot\vec{M})}{a}  \Bigg].
     \end{eqnarray}
From Eq. (\ref{aper2}), it can be observed that the $E_{i}$ are non-dynamic modes since no $\dot{E}_{i}$ terms are present. Hence, $E_{i}$'s are merely auxiliary fields, whose equations of motion give unique solutions to $E_{i}$ in terms of the dynamical modes $M_{i}$. To proceed, it is convenient to transform to 3-momentum space in order to get rid of the spatial derivatives. A further simplification is introduced by choosing the $z-$axis along the direction of 3-momentum $\vec{k}$, so that
\begin{eqnarray}
\label{bper0}
\dfrac{\partial f(t,\vec{x})}{\partial z} &=& -i\int d^{3}k k e^{-ikz}\tilde{f}(t,\vec{k}); \nonumber \\
\dfrac{\partial f(t,\vec{x})}{\partial x} &=& 0 = \dfrac{\partial f(t,\vec{x})}{\partial y} .
\end{eqnarray}
Substituting Eq. (\ref{bper0}) into Eq. (\ref{aper2}) yields,
\begin{eqnarray}
\label{aper3}
     S_{2}^{FT} = \int dt \ d^3k \ \Big[\frac{1}{2a} \left( \dot{\vec{\tilde{M}}}^{\dagger}\cdot\dot{\vec{\tilde{M}}} +ik(\dot{\tilde{M}}_x^{\dagger} \tilde{E}_y + h.c) - ik(\dot{\tilde{M}}_y^{\dagger} \tilde{E}_x + h.c ) + k^2(\tilde{E}_x^{\dagger}\tilde{E}_x + \tilde{E}_y^{\dagger}\tilde{E}_y) \right) -\frac{k^2}{2a^3}\tilde{M}_z^{\dagger}\tilde{M}_z \nonumber \\
     +   \left( \frac{m^2}{2} - \dfrac{(6\xi + 2\zeta)}{\kappa} \dot{H} - \dfrac{(12\xi + 3\zeta)}{\kappa} H^2 \right)a(\vec{\tilde{E}}^{\dagger}\cdot\vec{\tilde{E}}) - 
     \left( \frac{m^2}{2} - \dfrac{(6\xi + \zeta)}{\kappa} \dot{H} - \dfrac{(12\xi + 3\zeta)}{\kappa} H^2 \right)\dfrac{(\vec{\tilde{M}}^{\dagger}\cdot\vec{\tilde{M}})}{a}  \Big],
\end{eqnarray}
where, the notations are as follows: for any function $f$, $\tilde{f}\equiv \tilde{f}(t,\vec{k})$ and $\tilde{f}^{\dagger}\equiv \tilde{f}(t,-\vec{k})$. Now, varying the action with respect to $E_{x}^{\dagger}$, $E_y^{\dagger}$, $E_z^{\dagger}$, their equations of motion are found to be,
     \begin{eqnarray}
     \label{aper4}
     \tilde{E}_{x(y)} &=& +(-) \dfrac{ik\kappa\dot{\tilde{M}}_{y(x)}}{ \kappa [k^2+m^2 a^2] - [12\xi + 4\zeta] a^2\dot{H} - [24\xi + 6\zeta] a^2 H^2}; \nonumber \\
     \tilde{E}_{z} &=& 0.
     \end{eqnarray}
Substituting Eq. (\ref{aper4}) in Eq. (\ref{aper3}) yields an effective action, $S_{eff}$, with only quadratic kinetic terms, namely $\dot{\tilde{M}}_x^{\dagger} \dot{\tilde{M}}_x$, $\dot{\tilde{M}}_y^{\dagger} \dot{\tilde{M}}_y$ and $\dot{\tilde{M}}_z^{\dagger} \dot{\tilde{M}}_z$. The complete expression of $S_{eff}$ is not of present interest, except for the kinetic part which is given by,
\begin{eqnarray}
\label{bper1}
\left(S_{eff}\right)_{Kin.} = \int dt \ d^3k \left[\dfrac{N}{2a(N - \kappa k^2)} \dot{\tilde{M}}_x^{\dagger}\dot{\tilde{M}}_x + \dfrac{N}{2a(N - \kappa k^2)} \dot{\tilde{M}}_y^{\dagger} \dot{\tilde{M}}_y + \dfrac{1}{2a}\dot{\tilde{M}}_z^{\dagger} \dot{\tilde{M}}_z \right],
\end{eqnarray}
where, $N = \kappa(2k^2+m^2 a^2)- (12\xi + 4\zeta)a^2\dot{H} - (24\xi + 6\zeta)a^2 H^2$.

Clearly, Eq. (\ref{bper1}) implies that there is no ghost instability in the longitudinal mode $\tilde{M}_z$ whereas the coefficients of the remaining two transverse modes may come with a negative sign and hence give rise to instability. This possibility of ghosts is similar to that of vector inflation models, where the condition is reversed: the longitudinal mode causes instability while transverse modes are stable \cite{peloso2009}. We also note that for $\zeta = 0$ case, our result is in agreement with the conclusion of Ref. \cite{koivisto2009a}, and that instabilities exist for both isotropic and anisotropic spacetimes. Moreover, adding $R_{\mu\nu}$ coupling ($\zeta\neq 0$) does not help in treating the instability. If we demand that $S_{eff}$ be free of ghosts, then the following condition needs to be satisfied:
\begin{eqnarray}
\label{aper5}
\frac{k^2}{a^2}+m^2 > \frac{H^2}{\kappa} \left( (24\xi + 6\zeta) - (12\xi + 4\zeta)\epsilon \right).
\end{eqnarray}
In a special case of exact de-sitter space and taking $\zeta = 0$, the condition (\ref{aper5}) translates to, 
\begin{eqnarray}
\label{aper6}
k^2 > 4 a^2 H_0^2.
\end{eqnarray}
Eq.(\ref{aper6}) indicates that there will be no ghost in the action for sub-horizon modes only. While for  super-horizon modes the action will encounter ghost. This too is a familiar situation encountered in vector field models of inflation\cite{padhy2018}.

\section{\label{sec7}Conclusion}
We study the possibility of inflation with minimal and nonminimal models of rank-2 antisymmetric tensor fields. We find that the minimal model does not support inflation. Interesting features appear when a model with non-minimal coupling to gravity is considered, as a way to introduce a new parameter in the form of couplings $\xi$ and $\zeta$. It is possible to have solutions for de-Sitter space in nonminimal model that can support inflation. A simple bound on the couplings $\xi$ and $\zeta$ has been obtained from the de-Sitter solutions, and can support stable de-Sitter space under certain conditions. A detailed fixed point analysis will be carried out in future to ascertain the issue of stability of de-Sitter solutions. To study inflation, the slow roll analysis has been performed, and corresponding slow roll parameters $\epsilon$ and $\delta$ have been obtained. Validity of slow roll conditions has been checked. A notable feature of the present analysis is that the values of $\xi$, $\zeta$ and $\phi$ are sub-planckian in these models. 

The ghost instability analysis has been performed for perturbations to $B_{\mu\nu}$ (keeping the metric unperturbed). We find that while the longitudinal modes are ghost free, the transverse modes may admit ghosts. For a special case of exact de-Sitter space and $\zeta= 0$, only the sub-horizon modes are ghost free. It is noteworthy that the conditions encountered here are common in vector field models as well \cite{peloso2009,padhy2018}. 

The structure of Eqs. (\ref{bper1}) and (\ref{aper5}) hints towards the kind of modifications one would have to include in action (\ref{anon1}) to build a successful model of inflation with antisymmetric tensor field. An interesting possibility arises by adding a $U(1)$ symmetry breaking kinetic term to Eq. (\ref{anon1}): there are kinetic coulings between $E_{i}$ and $M_{i}$ modes, and any claim about instabilities cannot be made until one solves the coupled dynamical equations. This will be the subject of our subsequent study, and we speculate that possibly, instability problems could be resolved. In a future work, the full perturbation theory for such models may be developed, which will allow for phenomenologically relevant calculations. Possible extensions of this study include considering more combinations of coupling terms involving Ricci tensors and scalars, particularly $R^{2}$ coupling to tackle possible instabilities. Further studies may also involve the study of spontaneous Lorentz violation with antisymmetric fields in cosmological context and could provide significant insights for investigating signatures of new physics.

\begin{acknowledgments}
The manipulations in Sec. \ref{sec6} were done using Maple\texttrademark \footnote{Maple 2017.3, Maplesoft, a division of Waterloo Maple Inc., Waterloo, Ontario.} and cross-checked by hand. This work is partially supported by DST (Govt. of India) Grant No. SERB/PHY/2017041. The authors thank Tomi Koivisto for pointing out the results of Ref. \cite{koivisto2009a}.
\end{acknowledgments}

\bibliography{references,ref}
\end{document}